\documentclass[aps,pre,amsmath,amssymb,floatfix,twocolumn,10pt,groupedaddress,superscriptaddress]{revtex4-1}
\usepackage{graphicx,isomath}
\usepackage[colorlinks=true,linkcolor=blue,citecolor=blue,urlcolor=blue,pdfborderstyle={/S/U/W 1}]{hyperref}

\begin{document}

\title{Multifaceted Dynamics of Janus Oscillator Networks}
\author{Zachary G. Nicolaou}
\affiliation{Department of Physics and Astronomy, Northwestern University, Evanston, Illinois 60208, USA}
\author{Deniz Eroglu}
\thanks{Present address: Department of Bioinformatics and Genetics, Kadir Has University, 34083 Istanbul, Turkey}
\affiliation{Department of Physics and Astronomy, Northwestern University, Evanston, Illinois 60208, USA}
\author{Adilson E. Motter}
\affiliation{Department of Physics and Astronomy, Northwestern University, Evanston, Illinois 60208, USA}
\affiliation{Northwestern Institute on Complex Systems, Northwestern University, Evanston, Illinois 60208, USA}

\date{\today}
\begin{abstract}
Recent research has led to the discovery of fundamental new phenomena in network synchronization, including chimera states, explosive synchronization, and asymmetry-induced synchronization. Each of these phenomena has thus far been observed only in systems designed to exhibit that one phenomenon, which raises the questions of whether they are mutually compatible and, if so, under what conditions they co-occur. 
Here, we introduce a class of remarkably simple oscillator networks that concurrently exhibit all of these phenomena. 
The dynamical units consist of pairs of nonidentical phase oscillators, 
which we refer to as Janus oscillators by analogy with Janus particles and the mythological figure from which their name is derived. In contrast to previous studies, these networks exhibit (i) explosive synchronization 
with identical oscillators; (ii) extreme multistability of chimera states, including traveling, intermittent, and bouncing chimeras; and (iii) asymmetry-induced synchronization in which synchronization is promoted by random oscillator heterogeneity. These networks also exhibit the previously unobserved possibility of inverted synchronization transitions, in which a transition to a more synchronous state is induced by a reduction rather than an increase in the coupling strength. 
These various phenomena are shown to emerge under rather parsimonious conditions, and even in locally connected ring topologies, which has the potential to facilitate their use to control and manipulate synchronization in experiments. \\[1em]
DOI: \href{https://doi.org/10.1103/PhysRevX.9.011017}{10.1103/PhysRevX.9.011017}  \flushright \vspace{-2em} Subject Areas: Complex Systems, Nonlinear Dynamics \hspace{5.5em}
\end{abstract}

\maketitle
\section{Introduction}
 It has been the tradition of physics to describe complex behavior using 
 simple mathematical models. In such a description, of which the phenomenon of chaos offers many compelling examples \cite{ott2002}, 
 the behavioral complexity is emergent rather than explicitly coded in the model. In recent studies of network dynamics, the possibility of devising a complex network structure and nodal dynamics, and thus a complex model, has added a new dimension to this tradition. It has led to the discovery of fascinating new phenomena but has allowed for an easy departure from parsimonious models,  thus creating difficulties to  isolate the minimal requirements for the observed dynamical behavior. Among the new phenomena discovered, we highlight: (1) chimera states  \cite{kuramoto2002,abrams2004}, characterized by coexisting incoherence and synchrony in identically coupled identical oscillators; (2) explosive synchronization transitions \cite{gomezgardenes2011}, 
 in which the transition to synchronization becomes subcritical (hence abrupt) and hysteretic; and (3) asymmetry-induced synchronization (AIS) \cite{nishikawa2016,zhang2017asymmetry,zhang2018}, {a partial converse to the symmetry breaking exhibited by chimera states}, 
in which either the oscillators or their couplings need to be nonidentical for {synchronization} 
to prevail. These behaviors are unequivocally emergent, since they are not manifestly forged into the model. 
Yet, previous demonstrations of these various phenomena required the design of specific systems, in which the occurrence of the different types of behavior seemed to require different and sometimes complicated types of intrinsic dynamics, interaction structures, and coupling schemes.

\begin{figure}[t]
\includegraphics[width=\columnwidth]{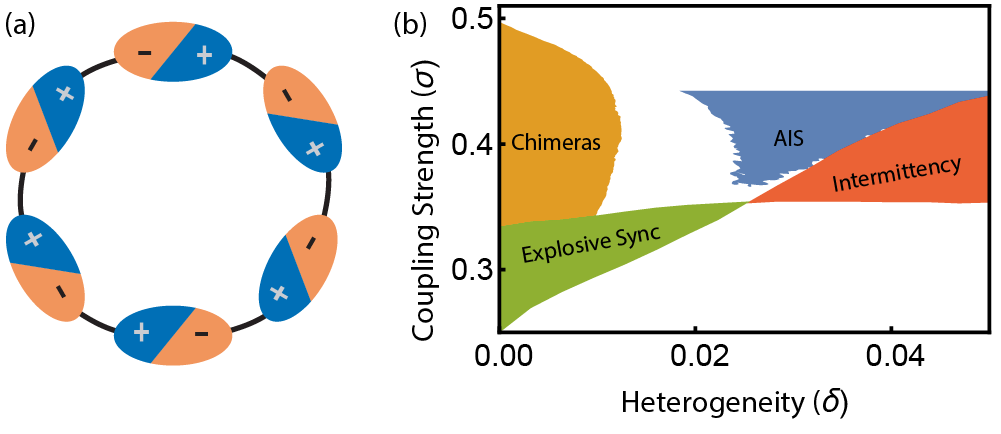}
\caption{ (a) Model network of phase-phase oscillators, in which phase oscillators with alternating frequencies are sine-coupled to their nearest neighbors on a ring. Regarding each pair of plus-minus oscillators as a 
Janus oscillator results in a symmetric system 
of oscillators.  
In such a ring (shown here with $N=6$ oscillators), any two Janus oscillators can be swapped by rotations without changing the overall configuration. (b) The various phenomena exhibited by a ring of $N=50$ Janus oscillators as the coupling strength as well as the  heterogeneity across the Janus oscillators are varied. Quantitative details about these regimes are presented in Sec.~\ref{hetsec}. }
\label{fig:ring}
\end{figure}

Here, we demonstrate the co-occurrence of chimera states, explosive synchronization, and a new form of AIS in a class of surprisingly simple oscillator networks. The dynamical units in these networks are two-dimensional phase-phase oscillators, which we term Janus oscillators by analogy with the  homonymous two-faced particles  (and the two-faced ancient Roman deity on which that name was based) \cite{1989_Casagrande}, 
since 
the components of the phase-phase pair are taken to have different natural frequencies.  
 Figure \ref{fig:ring} schematically shows an especially simple network we consider and the various dynamical behaviors it exhibits as a function of the coupling strength and oscillator heterogeneity. Importantly, our analysis of this new class of systems demonstrates for the first time the occurrence of (a) inverted synchronization transitions, (b) a plurality of chimeralike states, (c) explosive synchronization in the absence of correlations between the oscillator frequency and the network structure, and (d) synchronization induced by random oscillator heterogeneity. In particular, adding small oscillator heterogeneity destroys partially coherent states and makes fully phase-locked states more attractive, which represents a new form of AIS; and, for a certain range of coupling strengths,  a further increase in heterogeneity stabilizes a new 
type of dynamical behavior, which we call chimera intermittency.

Our model is partially inspired by the antiferromagnetic order first characterized in certain spin systems described below but that has since repeatedly reappeared in different physical systems as one of several universality classes of order. Janus oscillator models are an appropriate description of the oscillatory dynamics that can emerge in any driven system that exhibits antiferromagneticlike order in equilibrium. 

This paper proceeds with a description of the model in Sec.~\ref{model}, the presentation of the results in  Secs.~\ref{results1} and \ref{results2}, 
and a discussion of further implications 
in Sec.~\ref{concl}. Our presentation is complemented by an animated 
visualization of the main findings, which is included as Supplemental Material \cite{SM} and can be consulted before or after the text.

\section{Model Formulation}
\label{model}
\subsection{The model}

We define a Janus oscillator as a two-dimensional phase-phase oscillator in which each component has a distinct natural frequency.
We first consider ring networks of $n$  such oscillators, whose dynamics are governed by
\begin{eqnarray}
\dot \theta_{i}^1&  = \omega_{i}^1 + \beta\sin(\theta_{i}^2-\theta_{i}^1) +\sigma \sin(\theta_{i-1}^2-\theta_{i}^1), \label{eq:ode_1}\\
\dot \theta_{i}^2& = \omega_{i}^2 + \beta\sin(\theta_{i}^1-\theta_{i}^2)+\sigma\sin(\theta_{i+1}^1-\theta_{i}^2), \label{eq:ode_2}
\end{eqnarray}
for $i=0,\dots,n-1$, where subscripts indicate the Janus oscillator index while superscripts indicate the variable index. The parameters $\omega_{i}^{1}$ and $\omega_{i}^{2}$ are the natural frequencies of  oscillator $i$,  and the periodic boundary conditions are assured through the index convention $i= i \text{ mod } n$, where $\bmod$ is the modulo operation. Furthermore, the frequencies are assumed to be $\omega_{i}^{1} = \bar{\omega} - \omega/2$ and $\omega_{i}^{2} = \bar{\omega}  + \omega/2$, where $\bar{\omega} $ and $\omega$ are constants.
This assumption corresponds to the nearest-neighbor rotationally symmetric network of identical Janus oscillators $(\theta_{i}^{1}, \theta_{i}^{2})$ illustrated in Fig.~\ref{fig:ring}(a).
Finally, $\beta$ is the internal coupling strength between the phase-oscillator components of each Janus oscillator (i.e., in the same node) and $\sigma$ is the external coupling strength between oscillators in different nodes \cite{footnote1}.

The full parameter space of Eqs.~\eqref{eq:ode_1} and \eqref{eq:ode_2} is four dimensional, with coordinate axes $\bar{\omega} $, $\omega$, $\sigma$, and $\beta$.  However, this space can be substantially reduced without the loss of generality.  First, by changing to the corotating reference frame, we can eliminate   $\bar{\omega} $ so that each pair of oscillators has opposite natural frequencies. Second, by scaling the time by $1/\omega$ and $\sigma$ and $\beta$ by $\omega$, it is possible to set $\omega=1$.  Lastly, if either the $\sigma$ or $\beta$ are too large (i.e., greater than the critical value $\omega/2$ {at which the isolated pair of oscillators would synchronize}), the two phase components of each oscillator phase-lock and the dynamics can be reduced to a ring of single-phase oscillator nodes.  Thus, the dynamics of interest on the ring can be captured by fixing $\bar{\omega}=0$, $\omega=1$, and $\beta=\omega/4$, and varying {the external coupling strength between $0\leq \sigma \leq 0.6$ \cite{footnote2}}.

\subsection{Comparison with the existing models}
\begin{figure}[t]
\includegraphics[width=\columnwidth]{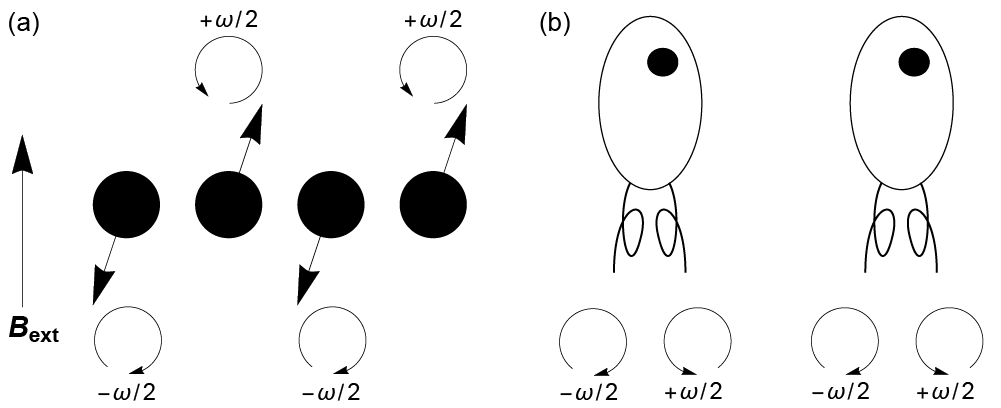}
\caption{Schematic representation of (a) the precession frequencies of dipoles that are dominantly aligned and antialigned with an applied magnetic field $\mathbfit{B}_{\mathrm{ext}}$ and (b) the \textit{Chlamydomonas} cells with counterrotating flagella.} 
\label{fig:spins}
\end{figure}

As noted above, our model is inspired in part by the image of an antiferromagnet driven by an external magnetic field. The ground state of an antiferromagnet consists of an alternating configuration of magnetic dipoles arranged on a lattice.  
 When an external magnetic field is applied to this lattice, the dipoles precess with alternating angular frequencies $\pm\omega/2$, as illustrated in Fig.~\ref{fig:spins}{(a)}.  The dynamics and synchronization of such coupled dipoles have recently found applications in spintronics. For example, the synchronization of arrays of spin-torque and spin-Hall nano-oscillators through coupling currents has been successfully described with the Kuramoto model (and with more complex models) and experimentally realized \cite{2017_Zaks, 2016:Flovik,2017:Awad,2014:Locatelli,footnote3}.  Our model could thus be applicable to an antiferromagnetic array of spin-torque or spin-Hall nano-oscillators. While still in development, the possibility of designing spintronic devices from antiferromagnets rather than ferromagnets has also recently attracted attention and increasingly detailed dynamical modeling \cite{2018_Baltz_Tserkovnyak,2018_Gomonay_Tserkovnyak}.  Another notable system that may be modeled with Janus oscillators is the counterrotating flagella in the cells of certain communities of algae called \textit{Chlamydomonas}.  The geometry of \textit{Chlamydomonas} cells is shown schematically in Fig.~\ref{fig:spins}(b). In communities of \textit{Chlamydomonas}, both internal cellular interactions and hydrodynamic interactions in the cellular environment are thought to induce local coupling between the phases describing each flagellum's position \cite{2012_Friedrich_Julicher,2016_Wan_Goldstein}. These flagella rotate with opposite natural frequencies, much like our Janus oscillators. Striking patterns of synchronization in communities of cells bearing single flagellum have been noted \cite{2009_Lauga}, and we propose that  communities of cells like \textit{Chlamydomonas} may rely on even more complex synchronization dynamics similar to the Janus oscillator dynamics we show here.

Previous studies on chimera states have also considered minimal models in order to gain a more complete understanding of this complex phenomenon. One such example is the solvable model of two populations of coupled oscillators \cite{abrams2008solvable}. Another example is the chimera identified in a globally coupled population of oscillators with delay feedback coupling \cite{Yeldesbay_2014}. Moreover, while initially found in nonlocally coupled oscillators, chimeras have also been noted in strictly locally coupled models of phase-amplitude oscillators in both networks \cite{2015_Laing,clerc2016chimera,2016_Hizanidis_Tsironis,2016_Bera_Ghosh,shepelev2017chimera} and continuous systems \cite{nicolaou2017chimera}.   Our phase-phase oscillator model constitutes a particularly simple totally symmetric system (i.e., it is both vertex transitive and edge transitive in the language of graph theory) which exhibits chimeras. 

Likewise, subcritical discontinuous synchronization transitions have already been known to occur in globally coupled networks of nonidentical inertial oscillators \cite{tanaka1997}, networks of phase oscillators with uniformly distributed natural frequencies \cite{pazo2005}, and networks of nonidentical delay-coupled oscillators \cite{yeung1999}. However, abrupt phase transitions in complex networks attracted a great deal of attention mainly after the discovery of explosive percolation on random \cite{achlioptas2009} and scale-free graphs \cite{cho2009}. Explosive synchronization was first observed numerically in scale-free networks \cite{gomezgardenes2011} and the discontinuity of the transition was later proven analytically for star networks \cite{vlasov2015}. Networks of globally coupled oscillators with bimodal frequency distributions have been shown to exhibit explosive synchronization transitions as well \cite{2009_Martens, 2016_Pietras}.  Details about the observations of subcritical transitions in globally coupled models are reviewed in Ref. \cite{2016_Boccaletti}. Here, in contrast with previous studies, explosive synchronization is shown to occur in a network of identical oscillators.  While previous work focused on explosive transitions from an asynchronous state to a single synchronous state, we explore an entire spectrum of transitions to a multitude of partially synchronous states.  Furthermore, the loss of this massive multistability with increasing heterogeneity leads to a new mechanism for asymmetry-induced synchronization.

In previous models, the traditional way to measure the degree of global synchronization of the whole population of oscillators has been the magnitude of Kuramoto order parameter 
\begin{equation}
r \equiv  \frac{1}{N} \sum_{j,k} \exp\left({\text{i}\theta_{j}^{k}}\right) ,
\end{equation}
which ranges from 0 
to 1 
with larger values typically corresponding to more synchronous states \cite{kuramoto1984,rodrigues2016}. However, the order parameter is not a perfect measure of the degree of synchronization. For example, the order parameter is $r=0$ for equidistantly distributed and phase-locked oscillators, and, thus, the order parameter of a partially coherent state may be higher than a fully phase-locked state. Therefore, the order parameter can be misleading for some cases and it is important to quantify such solutions with extra care.  Nevertheless, $r$ is a useful quantity to distinguish dynamical behaviors that we consider here. In addition, to distinguish cases in which $r$ does not adequately quantify synchronization, we introduce additional metrics below.

\begin{figure*}
\includegraphics[width=2\columnwidth]{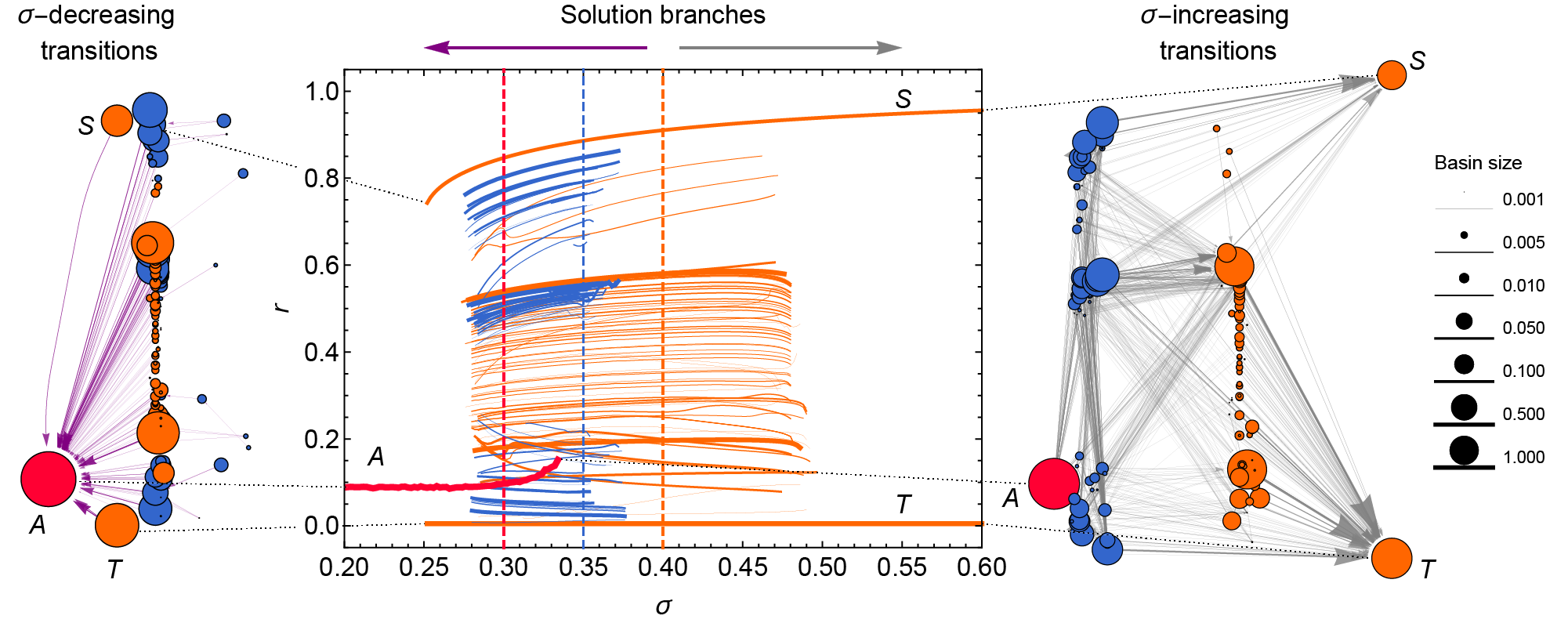}
\caption{Main panel:  Stable solution branches in the space defined by order parameter $r$ vs coupling strength $\sigma$. The line thickness indicates the size in the logarithmic scale of the corresponding  basins of attraction,  determined from $10^4$ random initial conditions at $\sigma=0.30$ (red lines), $\sigma=0.35$ (blue lines), and $\sigma=0.40$ (orange lines). Side panels: Networks of transitions between solution branches in the forward  (right) and backward (left) directions, generated by quasistatically increasing and decreasing $\sigma$, respectively. {The nodes represent solution branches, color coded as in the main panel, and with the radius 
representing the logarithmic size of the attraction basin, while the width of the directed links indicates the relative transition probability (i.e., the transition probabilities between branches times the logarithm basin size of the initial state). The node positions indicate the respective positions of left (right) ending points of solution branches in the left (right) panels, and reference asynchronous, synchronous, and twisted states are labeled with $A$, $S$, and $T$, respectively, with dotted lines showing the correspondence between the panels.} }
\label{fig:branches}
\end{figure*}

\section{The symmetric case}
\label{results1}

We start with the scenario described above, in which all Janus oscillators are identical and they are identically coupled through a ring network topology. Under such conditions, the system is symmetric in the sense that any two nodes can be swapped while leaving the equations of motion invariant, and fully synchronous states are the solutions that inherit that symmetry. This scenario is therefore a suitable model to investigate the interaction between symmetry breaking and synchronization phenomena.

\subsection{Numerical observations}
First, we show a global depiction of the solution branches and the transitions between those branches in the case of a ring of $n=50$ Janus oscillators with no heterogeneity in Fig.\ \ref{fig:branches}. To identify all solution branches, simulations with $10^4$ random initial conditions are performed for each value of $\sigma=0.30$, $\sigma=0.35$, and $\sigma=0.40$. These simulations wait for transients to die out before averaging the order parameter $r$ and the mean number of phase-locked oscillators $N_{\mathrm{locked}}$ \cite{source}.  
If all oscillators became phase locked, 
the state is deemed \textit{phase locked}; if some but not all of oscillators are phase locked, the state is deemed \textit{partially locked};  otherwise, the state is \textit{asynchronous}.  The final values of $r$ and $N_{\mathrm{locked}}$ are then binned to identify initial conditions that result in the same final state, and the number of initial conditions that end in each state is counted to estimate the size of the basin of attraction for each state. To map out the solution branches as a function of the coupling constant $\sigma$, simulations are performed that quasistatically vary $\sigma$ for each state identified from the random initial conditions. The network diagrams on the left and right in Fig.~\ref{fig:branches} show, respectively, the probability of transitions between various states under quasistatic changes in $\sigma$ (as determined through $10^3$ simulated transitions from each solution branch) in the decreasing and increasing directions, respectively. The nodes in these network diagrams represent the dynamical states with identical time-averaged values of both $r$ and $N_{\mathrm{locked}}$, corresponding to each solution branch we identify. They are arranged in space according to their critical coupling constants and order parameters.

\subsubsection{Solution branches}
\begin{figure}[t] 
\includegraphics[width=\columnwidth]{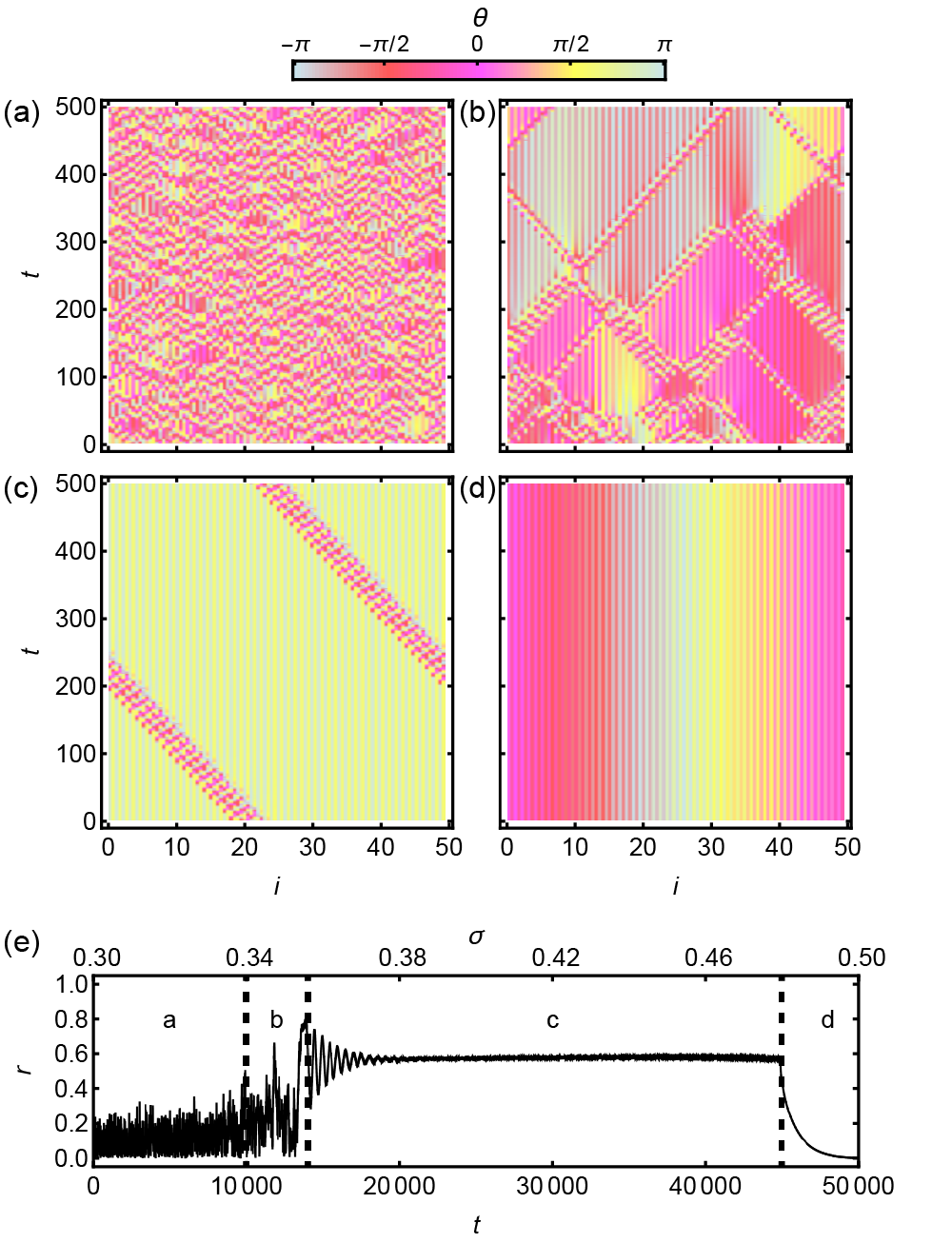}
\caption{Time evolution of various solution branches. (a) Asynchronous solutions, where each phase oscillator evolves independently with a mean frequency close to its natural frequency, are observed for small coupling constants. (b) Chimera intermittency, in which synchronous clusters intermittently appear and vanish, emerges as a transient from the asynchronous branch as the coupling constant increases past $\sigma_c^{\mathrm{async}} \approx 0.34$. 
(c) Chimera solutions, in which a cluster of asynchronous oscillators coexists with a cluster of phase-locked oscillators, 
are observed for intermediate coupling constants. 
(d) Phase-locked solutions, where all oscillators evolve with the same mean frequency, 
are observed for large coupling strengths.  (e) Order parameter $r$ vs the time for a state undergoing quasistatic coupling constant variation from $\sigma=0.3$ to $\sigma=0.5$.  During the periods between the dashed lines, the dynamics resembles each of the states in (a)-(d). Animations of these states are available in Supplemental Material \cite{SM}.}
\label{fig:timeplots}
\end{figure}

The time evolution of several states is shown in Fig.\ \ref{fig:timeplots}. 
After reaching a critical coupling, explosive synchronization occurs when the branch of the asynchronous state, shown in Fig.\ \ref{fig:timeplots}(a), disappears. The order parameter varies discontinuously at this critical coupling constant as the system jumps into another solution branch.  The loss of stability for the asynchronous branch occurs first through chimera intermittency, shown in Fig.\ \ref{fig:timeplots}(b). This intermittency is transient, and eventually the system settles into either a chimera state like the one in Fig.\ \ref{fig:timeplots}(c) or a fully phase-locked state like the twisted state shown in Fig.\ \ref{fig:timeplots}(d). As shown below,  the transient chimera intermittency can become stable and, thus, persistent when the system is not constrained to be symmetric.

The main mechanism generating the abrupt transition in explosive synchronization is the following. In scale-free networks, some oscillators in the system partially synchronize in clusters according to the similarities in their degrees and natural frequencies.  
These similarities can be regarded as a kind of symmetry between the nodes that synchronize.
Likewise, in networks of oscillators with symmetries, oscillators in the same symmetry orbit can synchronize even when there is no direct interaction between them. 
This kind of synchronization was first described as ``indirect synchronization'' \cite{1991_Okuda_Kuramoto}, but later dubbed \textit{remote synchronization} \cite{nicosia2013,zhang2017}. In our model, there exists such a remotely synchronized solution branch for arbitrarily small coupling constants, but this solution is neutrally stable in the linear stability analysis and is not attractive. This remotely synchronized state coexists with the asynchronous state, where no such clustering occurs. 
 However, the synchronized clusters in this state become phase locked with each other when the coupling increases past a critical value $\sigma_c^{\mathrm{sync}} = 0.25$, and at this point this synchronous state becomes stable and attractive. This phase-locking bifurcation is an example of a saddle-node bifurcation on the invariant circle \cite{1982_Aronson}, in which the center manifold of the saddle-node is the limit cycle corresponding to the remotely synchronized state. As the coupling constant quasistatically increases, eventually the asynchronous solution ceases to be stable at the critical coupling $\sigma_c^{\mathrm{async}} \approx 0.34$, and the system must move to either phase-locked cluster solutions or to other partially phase-locked solutions (see Fig.\ \ref{fig:branches}) during this explosive synchronization. The intermittent transient behavior shown in Fig.\ \ref{fig:timeplots}(b) occurs as $\sigma$ increases quasistatically past $\sigma_c^{\mathrm{async}}$. This behavior is reminiscent of spatiotemporal intermittency, which has been investigated, e.g., in the transition to turbulence \cite{chate1987transition}.

\begin{figure}[b]
\includegraphics[width=\columnwidth]{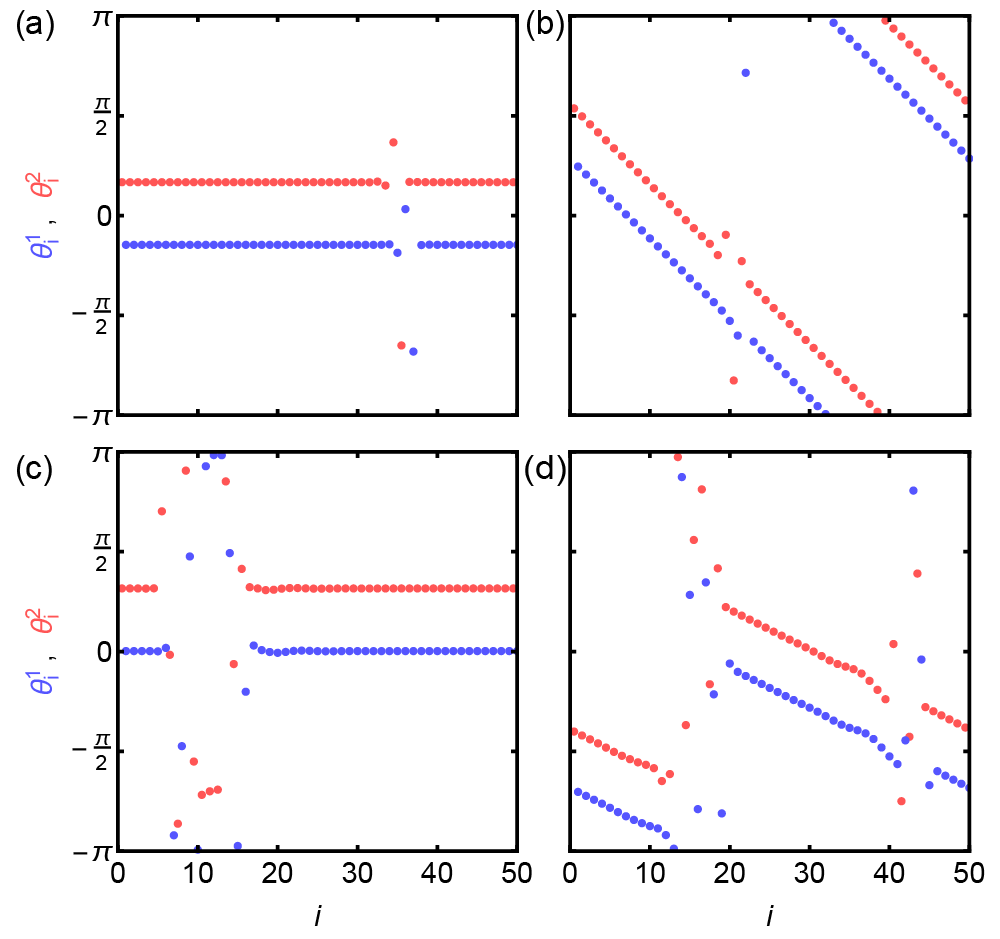}
\caption{Phase $\theta_{i}^{1,2}$ vs index $i$ for various partially locked chimera states.  Oscillators with natural frequency $-1/2$ are shown as blue dots, and oscillators with natural frequency $1/2$ are shown as red dots. In the chimera states in (a) and (b), most oscillators are phase locked, but a small number of oscillators drift around point defects.  Other chimera states are observed in (c) and (d) with one or more larger clusters of asynchronous oscillators. In all these cases, the structures are not fixed in space but move in time with an essentially constant velocity to the right or left. }
\label{fig:partial}
\end{figure}

In our simulations, we observe two kinds of phase-locked states: the twisted state with a small order parameter, where the phase increases through $2\pi$ around the ring depicted in Fig.\ \ref{fig:timeplots}\ (d), and the uniform phase state with a large order parameter, where all the oscillators have a similar phase, which is qualitatively similar. For intermediate coupling constants, several partially locked solution branches are observed with chimeralike collective behaviors, like that shown in Fig.\ \ref{fig:timeplots}\ (c).  Figure \ref{fig:partial} shows a selection of these chimera states in more detail.  These chimeras exist over differing intervals between $0.28 \lesssim  \sigma \lesssim 0.50$.
The chimeras in our system exhibit a symmetric behavior in which a domain of neighboring oscillators are phase locked and coexist with an asynchronous domain, as in standard chimeras, but these domains propagate to the left or right and visit each oscillator equally in time.
Different chimera states with dynamical boundaries have been observed in several other networks of coupled oscillators.  For example, the random meandering of chimera states has been previously noted as undesirable, and a control scheme has been proposed to affix them \cite{omelchenko2016tweezers}.  Another example is the breathing chimeras \cite{abrams2008solvable} in the solvable two-population model. Traveling chimera states have also been identified in phase and limit cycle oscillators with various nonlocal coupling schemes \cite{2014_Vulling,2014_Xie_Kao,2015_Hizanidis_Provata}.  The local coupling scheme in the ring of Janus oscillators constitutes a particularly simple model in which to study traveling chimera states.  

\subsubsection{Discontinuous and inverted transitions} 
A fundamental question associated with the observed multiplicity of stable solution branches concerns the {\it nature} of the transitions between branches as the coupling strength $\sigma$ is varied.
\begin{figure}[b]
\includegraphics[width=\columnwidth]{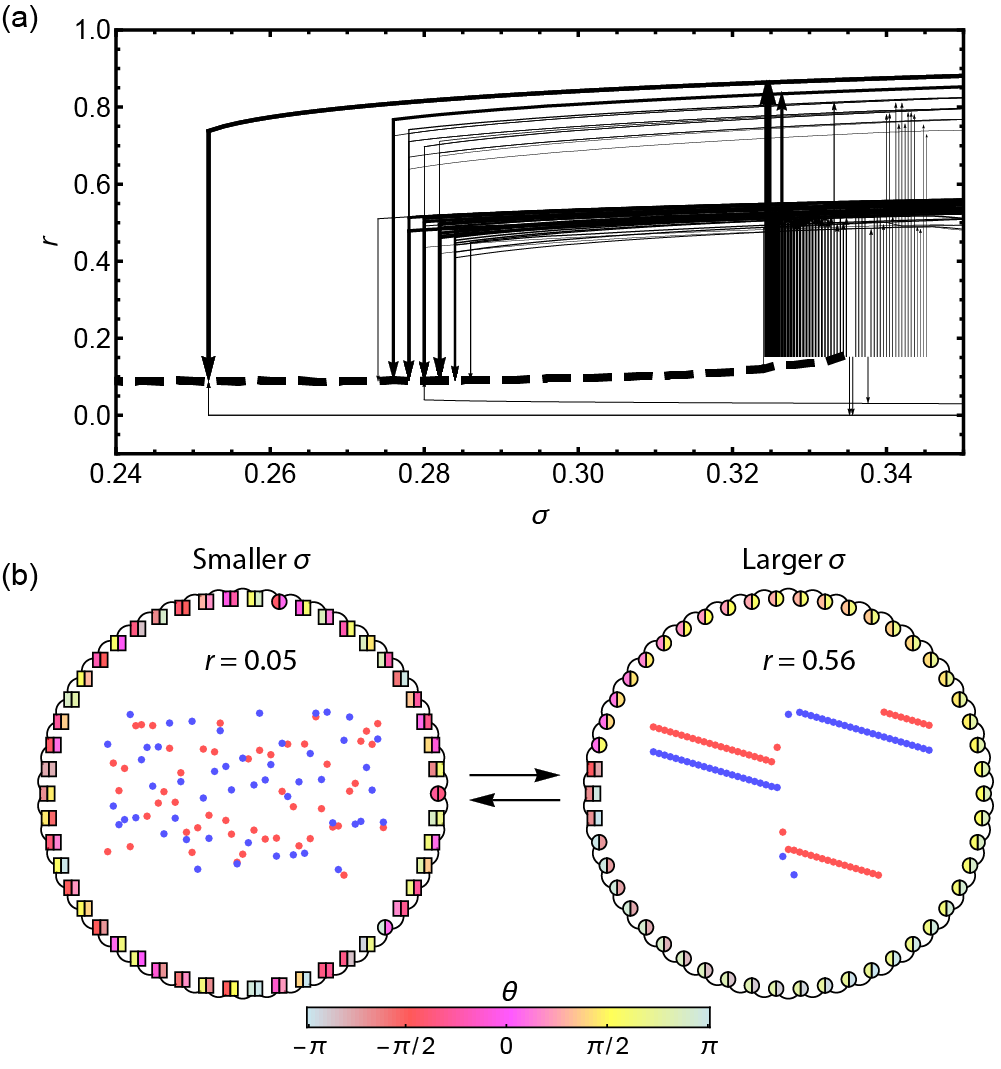}
\caption{Transitions to and from the asynchronous branch under quasistatic variations in the coupling strength. (a) Hysteresis curves around the asynchronous solution branch (thick dashed line), with transition probabilities indicated by the arrow thickness. (b) Example transitions to and from the asynchronous branch.  The circles (squares) depict phase-locked (asynchronous) Janus oscillators, and their colors show the instantaneous phases. An inset of the phases as plotted in Fig.~\ref{fig:partial} is shown in the center of the ring. }
\label{fig:transitions1}
\end{figure}
Among all the quasistatic transitions, those out of the asynchronous branch are the richest, with many possible final states. 
Figure~\ref{fig:transitions1}(a) shows the results for  detailed simulations of $10^4$ transitions from the asynchronous branch (dashed line) as the coupling strength
is quasistatically varied in the range $0.24\le\sigma\le0.35$.  
It follows that all solutions transition directly back to the asynchronous branch when the corresponding branches come to an end, 
 forming the various hysteresis loops shown in the figure.  A snapshot of the phases before and after one such transition is shown in Fig.\ \ref{fig:transitions1}(b).  
 As this example explicitly shows, the transitions from the asynchronous branch to phase-locked branches are discontinuous and  constitute
 genuine manifestations of explosive synchronization even though the network structure is a regular graph and the Janus oscillators are all identical. 
\begin{figure}[b]
\includegraphics[width=\columnwidth]{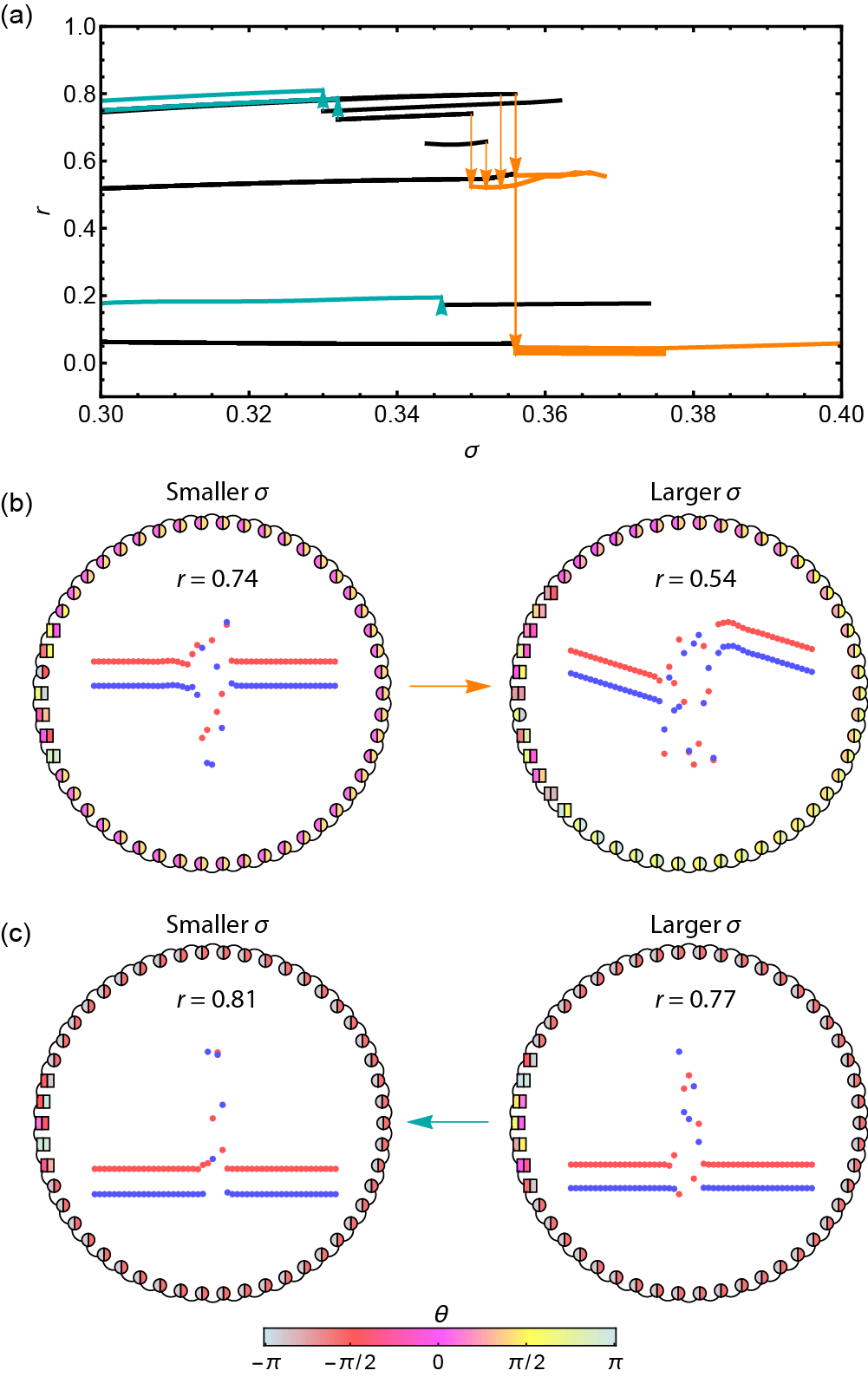}
\caption{Inverted transitions between solution branches under quasistatic variations in the coupling strength. (a) Inverted transitions for which an increase (or decrease) in the coupling strength results in a transition to a less (or more) synchronous solution. Example (b) forward- and (c) backward-inverted transitions, as in Fig.~\ref{fig:transitions1}(b).  
}  
\label{fig:transitions2}
\end{figure} 

A closer inspection of the simulations used to generate Fig.~\ref{fig:branches} shows that not all transitions occur in the expected direction, namely from less to more synchronous states as  $\sigma$ is increased and vice versa. Indeed, as shown in Fig.~\ref{fig:transitions2}(a) for simulations in the range $0.3\le\sigma\le0.4$, several transitions are \textit{inverted}. These are transitions between partially phase-locked solution branches in which an increase (or decrease) in the coupling strength results in a  decrease (or increase) in synchronization. Explicit examples of inverted transitions are shown in Fig.~\ref{fig:transitions2}(b) and Fig.~\ref{fig:transitions2}(c), where states for larger coupling strengths are visually less synchronous than those for smaller coupling strengths. All inverted transitions shown enjoy this defining characteristic and correspond to branches whose relative degree of synchronization is suitably measured by both the  mean number of phase-locked oscillators $N_{\mathrm{locked}}$ and the order parameter $r$. As in the case of explosive synchronization, inverted synchronization transitions are associated with subcritical bifurcations, and are thus discontinuous and hysteretic. We exclude from Fig.~\ref{fig:transitions2}(a) the transitions
 between the asynchronous branch and the bottom two twisted branches, which appear inverted with respect to  $r$, but this perception is in fact an artifact of the definition of $r$ noted above, 
which does not properly reflect the synchrony of the twisted state. 
For the forward-inverted transitions in Fig.~\ref{fig:transitions2}(b), a partially locked state becomes unstable under a quasistatic increase in the coupling constant, and chimera intermittency follows for a brief period.  This effectively randomizes the state, and there is some probability that, once the system settles into an attractor, the final state is less synchronous than the initial state.  On the other hand, in the backward-inverted transition in Fig.~\ref{fig:transitions2}(c), a small portion of the incoherent domain detaches from the rest of the incoherent domain on a quasistatic decrease in the couple constant and joins the coherent domain.   This process is not random but happens consistently under a quasistatic decrease in the coupling constant from this initial state.  The  animation in Supplemental Material is especially informative for understanding the dynamics of these inverted transitions \cite{SM}.
The inverted transitions characterized here  are analogous to negative compressibility transitions identified in mechanical metamaterials \cite{2012_Nicolaou_Motter}.

\subsection{Analytical results}
These rigorous numerical observations motivate us to pursue an analytical characterization 
of the observed patterns, 
which is possible by taking advantage of the symmetry in the model to identify a  large class of exact solution reductions.
These reductions are parametrized by the number of clusters,
 $2q$, representing the number of phase-locked groups of oscillators, and the twisting wave number $\nu$, representing the number of $2\pi$ phase windings each cluster undergoes around the ring. 
 We propose the following twisted-cluster ansatz:
\begin{align}
\theta_{i}^{1} &= \phi^1_{{i} \bmod q} + {i} \nu, \label{eq:ansatz1}\\
\theta_{i}^{2} &= \phi^2_{{i}\bmod q} + {i} \nu, \label{eq:ansatz2}
\end{align}
where $\bmod$ is, as before, the modulo operation. When Eqs.\ \eqref{eq:ansatz1} and \eqref{eq:ansatz2} are substituted into Eqs.\ \eqref{eq:ode_1} and \eqref{eq:ode_2}, the following reduction is derived:
\begin{align}
\dot{\phi^1_j} &= {\omega}/{2} + \beta \sin(\phi^2_j - \phi^1_j) + \sigma \sin(\phi^2_{j-1} - \phi^1_j - \nu), \label{eq:ode_3} \\
\dot{\phi^2_j}&= {-\omega}/{2} + \beta \sin(\phi^1_j - \phi^2_j)+ \sigma \sin(\phi^1_{j+1} - \phi^2_j + \nu), \label{eq:ode_4}
\end{align}
where $j=1,\ldots, q$ and $\phi^{1,2}_{0,q+1} \equiv \phi^{1,2}_{q,1}$.
 While this reduction may not appear any simpler on first glance, there are,  in fact, only $2q$ equations in Eqs.\ \eqref{eq:ode_3} and \eqref{eq:ode_4} as opposed to  the $N$ equations in Eqs.\ \eqref{eq:ode_1} and \eqref{eq:ode_2}. Accordingly, this result represents a significant dimension reduction for small $q$.

As an application of these exact reductions, consider the two-cluster solution for $q=1$.  For completely phase-locked two-cluster solutions, the difference  $\eta \equiv \phi^1_j - \phi^2_j$
is a constant.  It follows from Eqs.\ \eqref{eq:ode_3} and \eqref{eq:ode_4} that 
\begin{equation}
0 = \omega - 2\beta\sin\eta - 2\sigma \sin(\eta + \nu), 
\end{equation}
which has real solutions only when $\lvert \sigma \rvert \geq (\omega / 2-\beta)\cos\nu$.  The bifurcation that occurs as $\sigma$ is decreased below this value is the saddle-node bifurcation on the invariant circle discussed above. For $\nu=0$, $2\pi/N$, 
these phase-locking 
critical coupling constants are apparent in the leftmost transitions in Fig.\ \ref{fig:transitions1}(a).  
Similar bifurcations in the cases with higher $q$ and $\nu$ can be derived. However, these higher-order twisted-cluster solutions are not attractors or repellors but rather appear to be nonattracting invariant sets. The asynchronous domains in the chimera solutions in Fig.\ \ref{fig:partial} seem to exhibit an approximate twisted-cluster symmetry, as apparent in the visible pattern in the asynchronous domain in Fig.\ \ref{fig:timeplots}(c). 

\begin{figure}[b]
\includegraphics[width=\columnwidth]{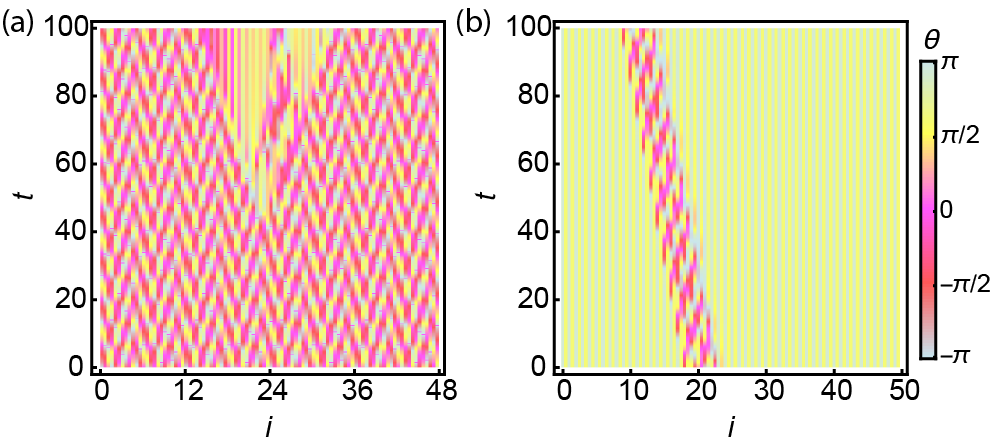}
\caption{(a) Time $t$ vs oscillator index $i$ for a perturbed cluster twisted solution with $q=2$ and $\nu=2\pi/12$  with a localized perturbation applied to the center node. (b) Time $t$ vs oscillator index $i$ for the chimera in Fig.~\ref{fig:timeplots}(c) with a finer timescale.   The propagating instability in (a) closely resembles the movement of the incoherent domain in (b). \label{fig:cluster}}
\end{figure}

Figure \ref{fig:cluster}(a) shows the evolution of a ring of Janus oscillators starting from an initial condition near one such unstable limit cycle solution to Eqs.~\eqref{eq:ode_3} and \eqref{eq:ode_4} with $q=2$ and $\nu=2\pi/12$. The ring is chosen with $N=48$ oscillators so an integer number of twists fits in the domain, and the initial condition is perturbed at the center to accelerate the decay of the unstable solution. The evolution consists of propagating fronts that closely resemble the patterns of the incoherent domains of a chimera, which is shown in Fig.~\ref{fig:cluster}(b) for comparison. While outside the scope of this work, it may be possible to interpret the chimera solutions as heteroclinic cycles connecting various cluster twisted solutions to phase-locked solutions \cite{2000_Buono}.  

Another possible analytic direction would be to employ the Ott-Antonsen reduction technique, which has been successful at describing many systems of globally coupled oscillators \cite{2015_Pikovsky}.  For example, (subcritical) explosive synchronization transitions in globally coupled networks with bimodal frequency distributions have been clearly characterized \cite{2009_Martens, 2016_Pietras}.  Extensions to nonlocally coupled networks have been proposed \cite{2009_Laing}, but challenges remain in applications to strictly local coupling schemes like the ring of Janus oscillators.

\subsection{General networks}
We conclude this section with a comment 
on more general symmetric networks of
Janus oscillators, described by 
\begin{equation}
\left[\begin{matrix} \dot \theta_{i}^{1}\\[1.1em]  \dot \theta_{i}^{2} \end{matrix}\right] = \left[\begin{matrix}  \omega_{i}^{1} \\[1.1em] \omega_{i}^{2}  \end{matrix}\right] +
\beta \! \left[\begin{matrix} \sin(\theta_{i}^{2}-\theta_{i}^{1})\\[1.1em]  \sin(\theta_{i}^{1}-\theta_{i}^{2})\end{matrix}\right] +
\sigma  \! \left[\begin{matrix} A_{ij} \sin(\theta_{j}^{2}-\theta_{i}^{1})\\[1.1em] A_{ji} \sin(\theta_{j}^{1}-\theta_{i}^{2})  \end{matrix}\right],
\end{equation}
where $A=(A_{ij})$ is an adjacency matrix, of which the system \eqref{eq:ode_1} and \eqref{eq:ode_2} is a special case. The entry $A_{ij}$ of this matrix is assumed to be
one when the first oscillator in node $i$ is coupled to the second oscillator of node $j$ and zero otherwise.   
Each link represents a  bidirectional coupling, but the adjacency matrix is not symmetric in general (and hence can be visualized as describing a directed interaction network)
as it also encodes 
which of the two oscillators in each node are connected. In this way, we consider networks with $k$ nearest-neighbor links on a $D$-dimensional lattice. Our numerical simulations reveal that $D=1$ rings with $k>1$ nearest neighbors also exhibit a large degree of multistability, including propagating chimeras. Thus, our core results appear to generalize to  such networks.
  On the other hand, for square lattice topologies in $D=2$ dimensions with $k=1$ nearest neighbors, numerical simulations show that while the explosive synchronization transitions persist, only the asynchronous solution and the fully phase-locked solutions are attractors.  
\newpage

 \section{Effects of heterogeneity}
 \label{results2}
We have seen that the simple system considered above exhibits rich dynamics, with many different solution branches. This was shown to be the case when the Janus oscillators are all identical and identically coupled, resulting in a globally symmetric system. However, in nature, the interacting elements are generally nonidentical and/or nonidentically coupled.  
Next, we consider the counterintuitive implications of  breaking the system symmetry by introducing oscillator heterogeneity or disorder in the network structure.

\subsection{Oscillator heterogeneity and AIS}
\label{hetsec}
\begin{figure}[b]
\includegraphics[width=\columnwidth]{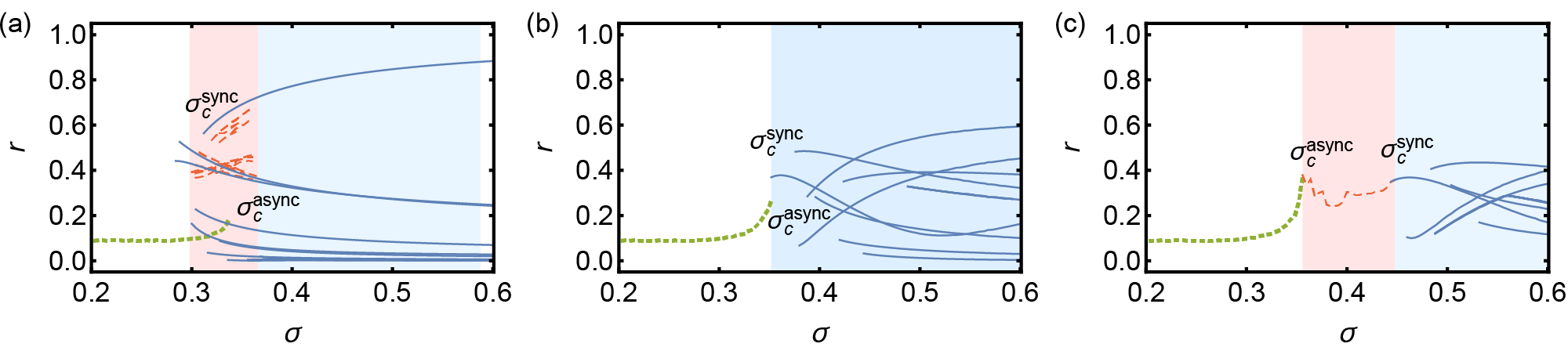} 
\caption{Impact of oscillator heterogeneity for one profile. Order parameter $r$ vs coupling strength $\sigma$ for (a) small heterogeneity ($\delta=0.005$), (b) moderate heterogeneity ($\delta=0.020$), and (c) large heterogeneity ($\delta=0.040$).
The lines indicate the asynchronous branch (green, dotted), fully phase-locked branches (blue, continuous), and  partially phase-locked branches (red, dashed).
 As the heterogeneity increases, the phase-locked solutions move to the right and the partially locked solutions disappear. 
The blue shade marks regions where only fully phase-locked solutions are present. The red shaded area in (a) marks the region where partially locked states exist and in (c) marks the region of stabilized chimera intermittency. \label{fig:hetero}}  
\end{figure}

 We consider ring networks as above but now 
for randomly perturbed oscillator frequencies: $\omega_{i}^{1,2} = \pm\omega/2+\delta p_{i}^{1,2}$, where  $p_{i}^{1,2}$ are   independent and identically distributed random variables drawn from a uniform distribution in $[-1/2,1/2],$ and the parameter $\delta$ defines the level of heterogeneity.
 As illustrated in Fig.\ \ref{fig:hetero} for one realization of the  {\it  heterogeneity profile} $p_{i}^{1,2}$, 
 the solution branches of the order parameter change in form as $\delta$ increases. One might have expected that introducing heterogeneity would decrease the degree of coherence  
 observed in this system.  However, this is not the case, as the partially phased-locked solutions in Fig.~\ref{fig:hetero}(a) disappear for increased heterogeneity, as shown  
 in Fig.~\ref{fig:hetero}(b) for $\delta=0.020$,
leaving the system with only coherent, phase-locked stable solutions for a range of coupling strengths. 
This result is evidence of AIS, a recently discovered effect \cite{nishikawa2016} in which oscillator heterogeneity can strengthen synchronization even when the oscillators are identically coupled. 
However, different from all previously reported cases of AIS \cite{nishikawa2016, zhang2017asymmetry, zhang2018}, in this scenario the phenomenon is  determined by {\it random} heterogeneity.
\onecolumngrid

\twocolumngrid
~
\pagebreak
\clearpage

\begin{figure}[b]
\includegraphics[width=\columnwidth]{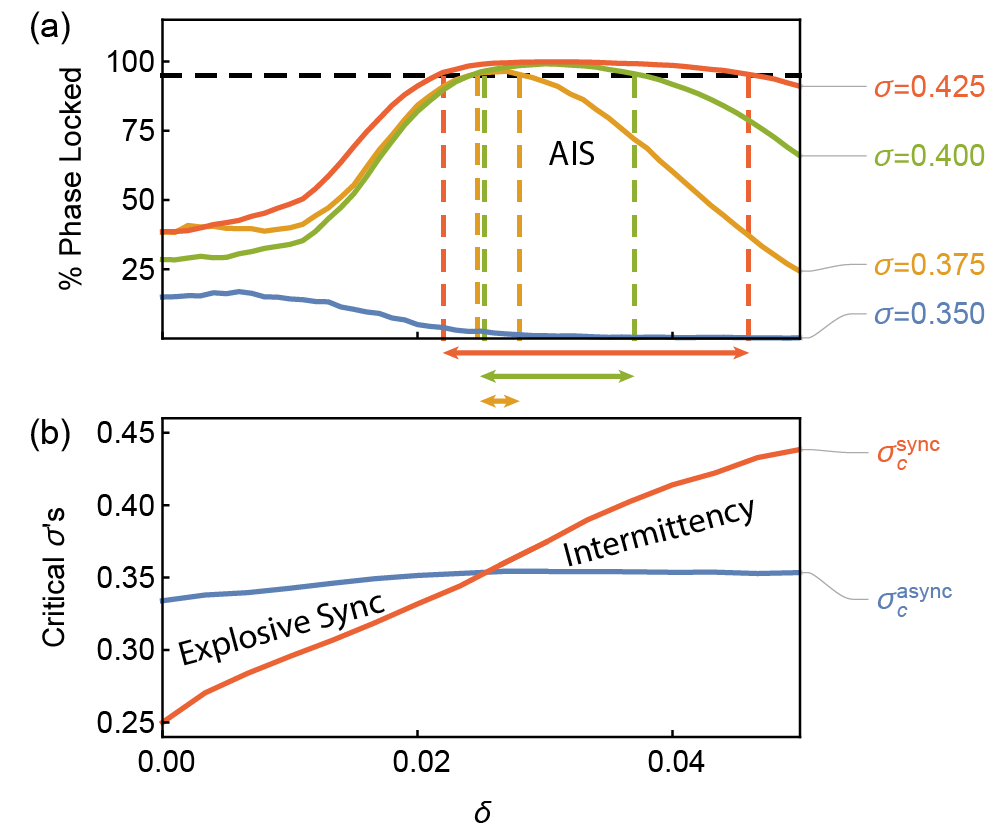} 
\caption{Statistical confirmation of AIS for moderate heterogeneity over many heterogeneity profiles. 
(a) Percentage of random initial conditions that become completely phase locked
and (b) 
critical coupling strengths as functions of $\delta$.  
The dashed lines and arrows show where the ring of Janus oscillators exhibits AIS given the $95\%$ threshold.
{Each value of $\delta$ is probed for $10^4$ different realizations of heterogeneity profiles and initial conditions to determine the percentages in (a) and $100$ heterogeneity profiles for each of $100$ different coupling constants to determine the critical values in (b).}}
\label{fig:ais1}
\end{figure}

We have seen that, for one heterogeneity profile, increasing the heterogeneity parameter $\delta$ can destroy the partially phase-locked states and cause the totally phase-locked states to dominate, but how does this trend hold statistically over many realizations of heterogeneity?  To address this question, we analyze the general effect of varying the heterogeneity profile.
Figure \ref{fig:ais1} shows results for $100$ random realizations of 
$p_{i}^{1,2}$ as $\delta$ is increased from $0$ to $0.05$.  We first examine the impact of heterogeneity on the size of the attraction basins of the various branches.  
As shown in Fig.\ \ref{fig:ais1}(a), a moderate level of heterogeneity can significantly increase the percentage of initial conditions that become completely ($100$\%) phase locked.
This increase is observed for intermediate coupling strengths, since the partially phase-locked solutions are destroyed in that region. {Since the AIS effect is quite prominent in this system, we can be rather strict and qualify the system as exhibiting AIS only when at least $95$\% of states (with differing random initial conditions and heterogeneity profiles) become completely phase locked. In addition, we require that, in the absence of heterogeneity (i.e., when $\delta=0$), a minority of states become phase locked for these coupling strengths, so that states which are already mostly synchronous in the absence of heterogeneity are excluded.
That is, for that range of coupling strengths, the introduction of random heterogeneity does not promote disorder. Instead, heterogeneity makes it overwhelmingly likely that the system will evolve into a  phase-locked state.
This result constitutes a striking example of AIS.}

The emergence of AIS in this system is intimately related with explosive synchronization and chimera intermittency.
This connection is best understood by considering the critical coupling strengths,  as shown in Fig.\ \ref{fig:ais1}(b).
For a small level of heterogeneity, the discontinuous  transitions resulting in explosive synchronization persist for all profile realizations, with the asynchronous branch disappearing at 
a larger coupling strength
 $\sigma=\sigma_c^{\mathrm{async}}$ than the coupling strength $\sigma_c^{\mathrm{sync}}$ at which the uniform, totally phase-locked synchronous branch appears. The other phase-locked solution branches appear at critical coupling constants that are slightly larger than $\sigma_c^{\mathrm{sync}}$, but all the critical coupling constants follow the trend of $\sigma_c^{\mathrm{sync}}$ as $\delta$ increases.  After increasing the level of heterogeneity past $\delta \approx 0.025$, where the average critical coupling strengths  $\sigma_c^{\mathrm{sync}}$  and $\sigma_c^{\mathrm{async}}$ coincide, 
the explosive synchronization disappears 
and the transition becomes continuous (i.e., supercritical). 
As $\delta$ is increased further, an irregular behavior sets in for a range of $\sigma$ 
in the asynchronous branch when {\it all} phase-locked branches exist only for coupling strengths larger than $\sigma_c^{\mathrm{async}}$.
This asynchronous state corresponds to the red-shaded region in Fig.\ \ref{fig:hetero}(c) and is, in fact, a stabilized and persistent form 
of the previously transient chimera intermittency 
 shown in Fig.~\ref{fig:timeplots}(b). 
Being  the only stable solution in this range of coupling parameters, this new  chimera state is
more accessible than the chimera states in Fig.\ \ref{fig:partial}. Thus, the occurrence of AIS in this system can be interpreted as follows: Introducing a small amount of heterogeneity into the model inhibits the explosive synchronization and eliminates partially phase-locked solutions,  which in turn results in a greater tendency toward phase locking.

To summarize the effects that oscillator heterogeneity has on the ring, we now describe the details about the various parameter regimes synthesized in Fig.~\ref{fig:ring}(b). The explosive synchronization regime (green) and intermittency regime (red) are determined from the critical coupling constants in Fig.~\ref{fig:ais1}(b). To quantify the chimera parameter regime (orange), simulations are performed to quasistatically increase $\delta$ and $\sigma$ starting from chimera initial conditions for $50$ different heterogeneity profiles. These simulations then detect when the chimeras cease existing in order to map the boundary.  The exact boundary where chimeras cease to exist depends on the profile, and the region in Fig.~\ref{fig:ring}(b) shows the average $\delta$ over the $50$ different profiles. To quantify the AIS parameter regime,  $1000$ simulations with different random initial conditions and heterogeneity profiles are performed for each point in the $\sigma$-$\delta$ plane. 
As in Fig.~\ref{fig:ais1}(a), the AIS region shows where the vast majority ($\ge95$\%) of random initial conditions and heterogeneity profiles evolve into completely (100\%) phase-locked states; the AIS region is delineated under the additional constraint that $< 50\%$ of the states are phase locked for $\delta=0$ in order to exclude parameters for which the homogeneous system is already mostly synchronized. The white region between the explosive synchronization, AIS, and chimera regions in Fig.~\ref{fig:ring}(b) represents areas where some heterogeneity profiles still exhibit attractive chimera states while the majority of heterogeneity profiles do not. This is a finite-size effect in which different behaviors will be observed depending on the exact profile of the oscillator heterogeneities.

\subsection{Network disorder and bouncing chimeras} 
\begin{figure}[b] 
\includegraphics[width=\columnwidth]{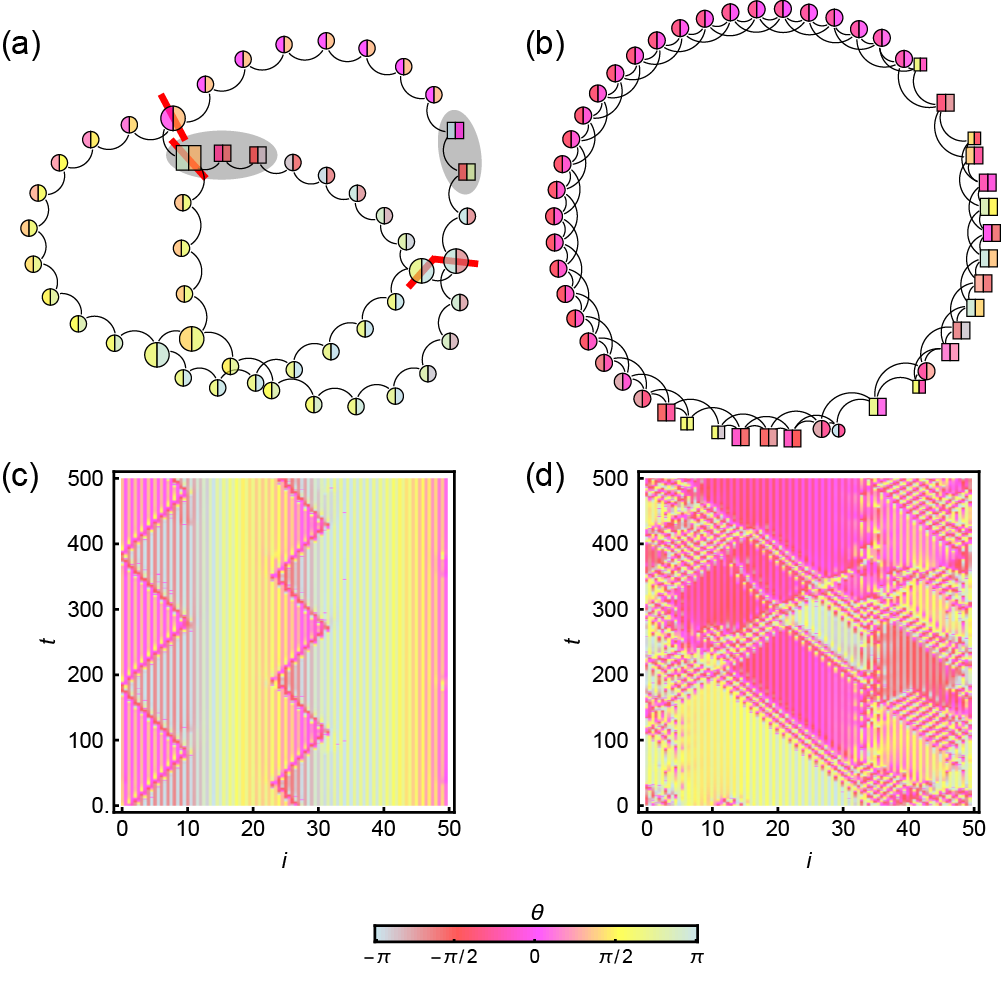} 
\caption{Effects caused by network disorder in rings with  [(a) and (c)] $k=1$ nearest neighbors  perturbed by  link addition and [(b) and (d)] $k=2$ nearest neighbors  perturbed by link removal.
The top panels show snapshots of the networks with the phases color coded on the oscillators, with squares showing instantaneously asynchronous oscillators, circles showing instantaneously phase-locked oscillators, and the node size reflecting the degree. The bottom panels show the corresponding time evolution 
of the phases. In (a),  the asynchronous domains (highlighted by the shaded ellipses) are confined within the domains marked by red lines.  }
\label{fig:smallworlds} 
\end{figure}
Network heterogeneity in the form of random link addition and/or removal is also studied on a ring with $k=1$ and $k=2$ nearest-neighbor links. Figure \ref{fig:smallworlds} shows typical results on such networks. When random links are added to the network, most chimeras are destroyed, but we observe that some chimera solutions persist and bounce off the network defects [Figs. \ref{fig:smallworlds}(a) and \ref{fig:smallworlds}(c)]. That is, these solutions  are defined by a moving group of asynchronous oscillators that  remain trapped between two nodes with added links, and they constitute what we term {\it bouncing} chimera states.  On the other hand, when links are randomly removed, the nodes with removed links do not phase-lock with their neighbors and remain unsynchronized for moderate coupling strengths [Figs. \ref{fig:smallworlds}(b) and \ref{fig:smallworlds}(d)]. Such nodes act as sources for traveling chimeras, resulting in chimera intermittency.

It follows that heterogeneity in the network structure can either inhibit synchronization---as one might expect---or reduce the degree of multistability by destroying most of the chimera states (leaving only the relatively rare bouncing chimeras)
and, thus, counterintuitively, promote synchronization. 
The latter can be interpreted as yet another form of AIS, where the symmetry of the system is broken by heterogeneity in the network structure rather than among the oscillators.  

\section{Concluding Remarks}
\label{concl}
Here, we introduced Janus oscillator networks as a new class of systems that can exhibit surprising behaviors even when the network structure itself is extremely simple. 
Our formulation builds on the principle that it is easier to understand complexity in simpler systems, provided that they can still include the relevant features. This principle has  been successfully explored in previous work that applied phase or dimension reductions~\cite{kuramoto1984,ott2008} 
and group theory~\cite{pecora2014}  
to the analysis of coupled oscillators. 
Owing to this simplicity, the prospects for experiments are very positive given that coupled phase oscillators, which are the components of Janus oscillator networks, have been implemented experimentally in electric \cite{Wiesenfeld:1996, Dorfler:2016}, chemical \cite{Kiss:2002, Miyazaki:2006}, and mechanical \cite{Pantaleone:2002, Mertens:2011} networks, among others \cite{Dorfler:2014}.

The significance of our study of Janus oscillator networks is multifold.~First, by demonstrating the co-occurrence of an array of behaviors previously observed in disparate systems, it shows how these different behaviors are related, and, thus, it will help to facilitate the manipulation of synchronization in applications. Second, it shows that these behaviors can be far more common than  anticipated and can even be unified in a simple model. 
Third, it allows the systematic characterization of transitions between an unprecedented number of stable branch states representing different levels of coherence, including various types of chimera states. Fourth, it reveals new kinds of behaviors, such as inverted synchronization transitions, characterized by a switch to a more  (less) synchronous state in response to a decrease (increase) in the coupling strength. Fifth, it demonstrates the generic occurrence of AIS in a remarkably simple system, which can facilitate the experimental demonstration of this phenomenon.
The latter is especially important given that this  counterintuitive phenomenon was discovered recently, is only starting to be understood, and is yet to be realized experimentally.

{Importantly, we uncovered unifying symmetry-based mechanisms behind the various phenomenon exhibited by Janus oscillator networks. The explosive synchronization in the symmetric ring follows because the synchronous solution branch bifurcates out of the symmetry-induced remotely synchronized solution, resulting in a subcritical bifurcation of the asynchronous solution branch.  We also suggested that the chimera states are heteroclinic cycles connecting the unstable cluster-twisted symmetry reductions and the synchronous solution. Introducing heterogeneity breaks the symmetry in the ring, inhibits explosive synchronization and chimera states, and promotes synchronous solutions, leading to AIS.}

This work can be expanded in many directions. On the one hand, while here we purposely focused on simple network structures, we speculate that an even richer range of behaviors  might be possible for Janus oscillators 
in complex network structures, which we expect will be explored by the community in future work. 
On the other hand, given that we focused on Janus oscillators with only two frequencies, it would be natural to also consider higher-dimensional Janus oscillators with three or more frequencies. 
Another extension would be to characterize  Janus oscillators whose ``faces'' are distinguished in terms of parameters other than the natural frequency (e.g., frustration parameter, delay, or oscillator type). In particular, experimentally accessible chemical oscillators \cite{Kiss:2002, Miyazaki:2006}, which are limit cycle oscillators of reactions in fluid cells, could be paired using cells with distinct geometries and chemical conditions.
Furthermore, it would be natural to consider Janus oscillators in time-varying networks to study oscillators that can move in space or coevolve with the network structure and analyze the resulting interactions between swarming, self-organization, and synchronization \cite{keeffe2017}. 
Finally,  we expect that this line of theoretical work will stimulate the experimental study of Janus oscillator networks and lead to yet new discoveries in the lab.
\vspace{-1em}
\section*{Acknowledgments}
This work was supported by U.S. Army Research Office Grant No. W911NF-15-1-0272.


\begin{thebibliography}{99}
\bibitem{footnote2}{We pad this interval beyond $\omega/2$ to numerically demonstrate that the dynamics do becomes fully synchronous at this critical value for the isolated pair of oscillators. }
\end{thebibliography}
\end{document}